# Sunbeam: Near-Sun Statites as Beam Platforms for Beam-Driven Rockets

Jeffrey K. Greason[a,b,*] Gerrit Bruhaug[c]

[a] Tau Zero Foundation, Post Office Box 1521, Broomfield, CO, 80038, USA
[b] Electric Sky, Midland, TX 79707 USA

[c] Los Alamos National Laboratory, Los Alamos, NM 87545, USA

Abstract: We outline a method of beamed power for propulsion that utilizes relativistic electron beams. The physics of charged particle beam propagation in the space plasma environment is discussed and the long-range (> 100 AU) advantage of relativistic electron beams is emphasized. A preliminary statite-based beam emitter for powering probes to ≈0.1$c$ is proposed and the challenges in beamed-power uses are explored.

## 1. Introduction & Background

Reaching velocities useful for practical interstellar flight is fundamentally a problem of supplying sufficient kinetic energy to the spacecraft. A 1000-kg interstellar probe (comparable to the mass of the Voyager spacecraft) at 20% of the speed of light carries 1.8x10$^{18}$ J of kinetic energy. Due to the limitations of energy sources carried aboard the vehicle, transmitting energy or momentum to the vehicle from external sources (beamriders) have long been recognized as one of the promising approaches for interstellar flight. Concepts include pushing with the direct photon pressure of laser beams [1] or microwaves [2] or via the pressure of macroparticles [3], [4], [5] or elementary particles [6]. The challenge of beamrider concepts of all types is that the capital cost of the beam source is very high; the power requirements for photon pressure are inherently high, and the limited range of the beam requires high acceleration to acquire the interstellar cruise velocity before leaving the range of the beam [7]. A current project, Breakthrough Starshot [7], envisions beam ranges on the order of 0.1 AU from a very large laser array that directly pushes on light sails. Using the energy of the beam to expel reaction mass from the vehicle requires less power for a given thrust than using the beam momentum directly for missions where the total velocity change ($\Delta v$ is much less than the beam velocity [8], [9]), so long as exhaust velocities of roughly one third to one half the $\Delta v$ are employed. However, neither mode of operation fundamentally changes the tradeoff requirements for beams.

Increasing the range of the beam is a powerful approach for reducing the power and capital cost of the beam transmitter. For acceleration limited by power, in the ideal case of perfect efficiency and no upper limit to acceleration the range of a given beam is given by Eq (1):

$$d_{\text{beam}} = \frac{1}{3} v_{\text{final}}^3 \frac{M_{\text{ship}}}{P_{\text{ideal}}} \qquad (1).$$

It should be noted that $d_{beam}$ is the distance of the propulsive manuever. So, for a given final velocity, the required power for the beam scales inversely with beam range. It is therefore worth seeking beams where practical transmitter hardware can reach dramatically longer ranges on the order of 100 to 1000 AU (thousands of times longer than envisioned by Breakthrough Starshot).

Keeping a beam on target for such long distances involves long times as well, which is how a given energy can be achieved at lower power. Rather than minutes, a ship using such a beam would be in the beam for weeks or months. That suggests a space-based rather than a ground-based beam, and for practical interstellar probes, the beam power is likely to be in the range of a gigawatt or more. Therefore, the ideal beam architecture would not only permit long ranges but lend itself to harvesting the power from space in a manner using a minimum of space-based infrastructure.

## 2. Physics of Long-Range Charged-Particle Beams

The physics of electromagnetic beams for power and momentum transfer have been well surveyed in the interstellar literature [1] [2] [8] [6] and to reach 100 AU or greater distances, the relevant beam transmitter and focusing optics, while not beyond the bounds of imagination, represent large scale space infrastructure beyond our near-term expectations. This leaves open particle or macroparticle beams. Macroparticle beams can be long range and have considerable promise [3], [4], [5] but high speed macroparticle beams of high average power are expensive, and affordable strategies [10] have thus far been limited to 2% of *c*.

For elementary particle beams, neutral particle beams [11] offer promise, though achieving the low divergence needed for long range is challenging [12], but the discussions in the literature have generally dismissed charged particle beams for long-range use because of space charge limitations. The mutual electrostatic repulsion of the identically charged particles in the beam is expected to cause the beam to expand far too rapidly for long range use. This is in spite of the fact that rather well-collimated charged particle beams are regularly observed in the space environment (for example, in solar flares, in the sheet-pinched heliospheric current that gives rise to the interplanetary magnetic field, in interstellar filaments which stretch for tens of light years [13] and in astrophysical jets stretching across 100,000 light years [14]). The propagation of charged particle beams is a well-established field [15] and has also shown well collimated propagation of charged beams in vacuum, rarified gases, and atmospheric pressure gases.

a. *Relativistically Confined Beams*

In a true vacuum, or in a space plasma too diffuse to affect the propagation of an electron beam, the mutual repulsion between charged electrons acts to expand the beam via space charge and have thus been dismissed in some past studies focused on neutral particle beams [12]. However, the relativistic effects of a highly relativistic beam, with electrons being the easiest to accelerate, can greatly reduce the effects of space charge. Relativistic ion beams may also be possible over relevant ranges [16] but as will be seen, the physics of beam propagation favor a high Lorentz factor ($\gamma \gg 1$) and ion beams have other complications associated with higher beam current, so we will confine the analysis in this paper to relativistic electron beams. It is quite easy to accelerate electron beams close to the speed of light ($\gamma \gg 1$). This leads to a form of beam confinement well known in the accelerator and beam physics communities, sometimes called the *relativistic pinch*. In the frame of reference co-moving with the electron beam, this can be thought of as relativistic time dilation; there is not enough time in the beam frame for the space charge to spread the beam very far.  Or, in a frame of reference at rest with respect to the Sun, one can think of these as relativistic increases in the electron momentum effectively reducing the charge-to-mass ratio of the electrons, again slowing the spread of the beam.   For a relativistic electron beam without compensating charges, the characteristic *doubling distance* $z_d$ in which the beam doubles in diameter due to space-charge repulsion is, in the relativistic limit ($\beta \approx 1$): [15]

$$z_\text{d} = \sqrt{\frac{I_0 \gamma}{|I_\text{b}|}} (\gamma^2 - 1)^{\frac{3}{4}} r_0 \qquad (2).$$

where, $I_\text{b}$ is the beam current, $\gamma$ is the Lorentz factor, and $r_0$ is the beam radius at the narrow portion of a diverging beam.  $I_0$, the *characteristic current* is, for electrons:

$$I_0 = \frac{4\pi \varepsilon_0 m_e c^3}{q} \approx 17000\text{A} \qquad (3).$$

Since the kinetic energy of an electron at relativistic velocity is:

$$E_\text{k} = (\gamma - 1) m_e c^2 \qquad (4).$$

For large $\gamma$, the electron beam is primarily a carrier of energy, rather than momentum.  The beam power is then:

$$P = (\gamma - 1) \varphi_\text{e} m_e c^2 \qquad (5)$$

where $\varphi_\text{e}$ is the number flux of electrons per second.  The beam current, $I_\text{b}$, is then simply:

$$I_\text{b} = \varphi_\text{e} q \qquad (6).$$

where $q$ is the elementary charge. We can then rewrite beam power as:

$$P = (\gamma - 1) \frac{I_\text{b}}{q} m_e c^2 \qquad (7).$$

Rearranging to express $I_\text{b}$ in terms of $P$:

$$I_\text{b} = \frac{Pq}{(\gamma - 1) m_e c^2} \qquad (8)$$

and substituting into the expression for $z_d$ gives:

$$z_\text{d} = \sqrt{\frac{I_0 \gamma (\gamma - 1) m_e c^2}{Pq}} (\gamma^2 - 1)^{\frac{3}{4}} r_0 \qquad (9)$$

tor $\gamma \gg 1$, this simplifies to:

$$z_\text{d} \cong \sqrt{\frac{I_0 m_e c^2}{Pq}} \gamma^{\frac{5}{2}} r_0 \qquad (10).$$

Higher power shortens the doubling distance, but this can be overcome by increasing the beam $\gamma$ which has extremely favorable scaling. Doubling the beam $\gamma$ allows for a 32-fold increase in power for the same beam doubling distance.

To work out the required electron energy, we rearrange the equation:

$$\gamma^{\frac{5}{2}} \cong \frac{z_d}{r_0} \sqrt{\frac{Pq}{I_0 m_e c^2}} \quad (10).$$

which clearly shows the desirability of a large transmitting and receiving aperture (just as with electromagnetic beams).  However, there is no special difficulty in doing so; a stream of adjacent electron beams without substantial neutralization from space plasma in between the streams can be treated for space charge purposes as a single beam.  Suppose the radius at the transmitter of the 1 GW beam is 10 m, and the receiver, 40 m, at a distance greater than 100 AU, then the doubling distance is 50 AU, and the required $\gamma$ is 36,500, or a beam energy of ≈19 GeV.  While different parameters could be chosen, the conclusion would remain the same; holding an electron beam together over the distances we seek for interstellar acceleration requires very high electron energy.

b. *Beams pinched in the background space plasma*

Charged particle beams propagating through the interplanetary and interstellar medium interact with the background space plasma.  Once the beam is established, electrons in the background plasma move towards a positive ion beam or away from a negative electron beam.  The beam, in its own reference frame, sees this locally non-neutral plasma as a current which acts to confine the beam, offset by the space charge of the beam and its internal pressure (from the effects of finite beam lateral temperature, which is typically described as beam emittance).

Relativistic beams are easier to pinch; they must satisfy the Budker self-focusing condition [15]:

$$1 > f_e > \frac{1}{\gamma^2} \quad (11)$$

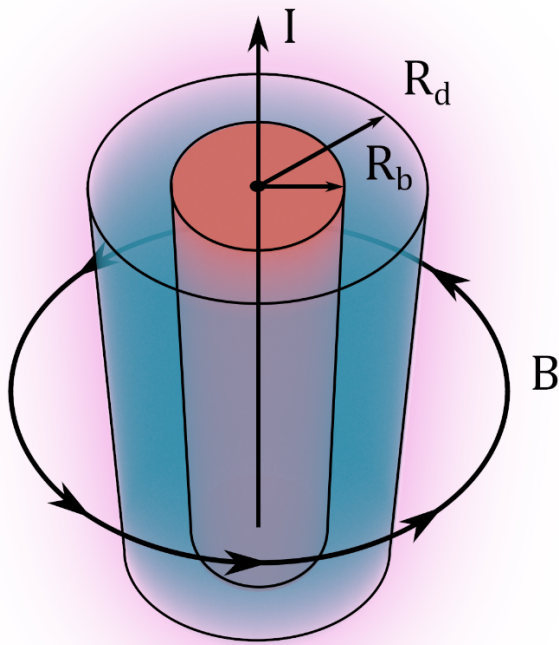

*Figure 1: Pinched beam in low density plasma with depletion region*

where $f_e$ is the ratio of positive charge background plasma number density to the negative electron charge number density in the beam. In a diffuse space plasma, the ion number density of the space plasma is much less than the electron number density in the beam, the *under-dense plasma* or *ion-focused regime* [17]. In that regime, in the reference frame of the background plasma, the concentration of negative electron charges will deplete the plasma of background plasma electrons not only throughout the beam cross section, but to a thickness beyond the beam, the depletion layer thickness varying roughly with the voltage to the 3/2 power [18], and the resulting current producing a circumferential magnetic field to pinch the beam, as illustrated in Figure 1. In the reference frame of the beam, only currents from particles inside the beam radius can contribute to pinching the beam, so assuming that the ions are approximately stationary during the time of one beam bunch propagating through the plasma, we reach the classic equation for the radius of a self-pinched beam in equilibrium (beam radius constant) [15]:

$$r_{\text{pinch}} = \frac{\epsilon_{\text{eff}}}{\sqrt{|K_{\text{gen}}|}} \qquad (12)$$

where $\epsilon_{\text{eff}}$ is the effective emittance of the beam (accounting for both beam lateral temperature and background ion temperature effects [19]), and $K_{\text{gen}}$ is the generalized perveance [15] in the relativistic limit.

$$K_{\text{gen}} = \frac{I_b}{I_0} \frac{2}{\gamma^3} (1 - \gamma^2 f_e) \qquad (13).$$

While for unpinched propagation, emittance is a function of the beam quality at the source, in the pinched limit, the background ion temperature contributes [15]:

$$\epsilon_{\text{eff}} = \sqrt{\epsilon^2 + \frac{4 r_{\text{pinch}}^2 k_B T_i}{\gamma m_e c^2}} \qquad \{14\}.$$

Under the conditions of underdense plasma, $f_e$ is itself a function of beam radius. Assuming the positive charge density in the plasma equals the number density of ions $n_{\text{ip}}$:

$$f_e = \frac{n_{\text{ip}} q c \pi r_{\text{pinch}}^2}{I_b} \qquad (15).$$

The pinch radius is then that radius at which both Eqs. (12) and (15) are satisfied, using the definitions of Eqs. (13) and (14).

As $\gamma^2 f_e$ approaches unity, the pinched beam radius blows up, so if we confine ourselves to the case $\gamma^2 f_e \gg 1$ we can combine these equations and simplify. If we further restrict ourselves to $\gamma \gg 1$ and to the case we are likely to encounter where the beam-quality driven emittance of the beam is negligibly small compared to the effect of background ion temperature in Eq. (14), we can write the simple and somewhat counterintuitive result:

$$r_p = \sqrt{\frac{2 I_0 k_B T_i}{\pi m_e c^3 n_{\text{ip}} q}} \qquad (16)$$

where $r_p$ is the approximate beam pinch radius within the domain covered by these approximations. In that regime, the factors of $\gamma$ and $I_b$ and $\epsilon$ drop out leaving only a dependence on background plasma density $n_{ip}$ and $T_i$, both of which are background plasma properties. Physically, this is because a higher radius increases the effective emittance due to background ion temperature (an effect which only applies to pinched beams) and increases $f_e$ in such a manner that the effects cancel. The table below shows examples of pinch radius for various conditions in the solar system, bearing in mind that real space plasmas vary over time and these values are only approximate.

| Distance from Sun Center (AU) | Plasma density (m$^{-3}$) | Ion Temperature (K) | Pinch Radius (m) | References |
|---|---|---|---|---|
| 0.04 | 6x10$^{10}$ | 700,000 | 0.33 | [20] |
| 1 | 4x10$^6$ | 86,500 | 14 | [21] |
| 115 | 2x10$^3$ | 2,000 | 98 | [22] |
| >140 (ISM) | 1x10$^5$ | 7,000 | 26 | [23], [24] |

Table 1: Distance from the sun, plasma density at that distance, ion temperature, resulting beam pinch radius, and references for space plasma parameters.

The pinched beam gets to rather large radii in the outer solar system just before the heliopause due to low plasma density, then reduces again in the interstellar medium. If the larger radius can be accommodated by the receiver via a focusing field, the pinched beam may be a suitable solution throughout; at higher $\gamma$ it may be more practical to use the relativistically confined beam out past the heliopause and then allow it to pinch in the interstellar medium (ISM).

c. *Beam Stability*

Electron beams propagating through a background space plasma are subject to both energy loss mechanisms and various instabilities, both of which can cause the beam to break up. Collisions with the background ions can scatter electrons out of the beam and eventually cause beam breakup. The characteristic equation (Nordsieck equation) for electron beams is equation 4.74 in [25] and for the case where the distance involved is very small compared to the radiation length of the medium, can be written in terms of a doubling length from Nordsieck effects:

$$z_{dN} = \frac{2\ln(2)X_0}{1+\frac{P_N}{P_b}} \tag{17}$$

where $P_b$ is the beam power, and $P_N$ is a constant, the Nordsieck power, a relationship between natural constants that defines a power scale for beams [25], which evaluates to 15 TW, and $X_0$ is the radiation length for the medium (slightly rewritten for a hydrogen plasma from equation 4.66 in [25]):

$$\frac{1}{X_0} = 4\alpha r_e^2 n_{ip} \ln(183) \tag{18}$$

where $\alpha$ is the fine structure constant and $r_e$ the classical electron radius. For the plasma properties in Table 1, the result is that the radius doubling length even at 0.04 AU from the sun for a 1 GW beam is over 8500 AU, by the time we reach 1 AU from the sun, the doubling length is over 10$^8$ AU, and for the conditions in the local interstellar medium, the doubling length is over 5x10$^9$ AU. Therefore, energy loss from scattering in the background plasma does not appear significant (and will become less so for higher power beams).

The high velocity electron beam passing through the essentially stationary background plasma ions is an interaction between two charged particle streams and as such is subject to the two-stream instability. Note that even when stable, the ions will develop a repeating density variation over timescales set by the ion plasma frequency, and that will in turn give rise to a velocity variation in the electron beam [27]. However, if the beam current exceeds a critical current threshold, these oscillations will become unstable and break up the beam. The region of stable beam operation is, in the relativistic limit (β≈1): [28], [27]

$$I_b \ll (425 \text{ A}) \frac{\gamma^3}{\left[1 + \left(f_e \gamma^3 \frac{m_e}{m_i}\right)^{\frac{1}{3}}\right]^3}$$

For a 1 GW electron beam, with $\gamma$ of 100, beam current is 19.8 A. Under the conditions given previously for an equilibrium pinch, $f_e$ is in the range of $4 \times 10^{-4}$. The critical current for two stream instability onset is then ≈400 MA, or over 20 million times above the relevant beam current, so it does not appear that two-stream instability should interfere with beam propagation in the expected domain.

Ion-focused beam propagation can also have a *hose* instability in which the beam will move off-center due to interactions with the background plasma electrons expelled from the channel. While not discussed in detail here, this is known theoretically to be suppressed by complete expulsion of background plasma electrons from the channel [29] and is observed experimentally to be a feature experienced in high-density gases and suppressed in low-density gases [30], so we do not expect this instability to limit beam propagation in the expected regime in the steady state (as discussed below, this is a potential difficulty in starting the beam).

There are of course many other possible instabilities; for a variety of reasons, we anticipate that an amplitude-modulated or *bunched* beam will be employed, and so long-period instabilities suppressed by limiting the duration of a single bunch. However, a further analysis and particle-in-cell simulations would be helpful in assessing the feasibility of this strategy for beaming power to spacecraft over extremely long ranges.

d. *Effect of Background Magnetic Field*

The interaction of the relativistic electron beam with the magnetic fields present in space, especially the interplanetary magnetic field, is not well understood at this time and may pose a serious obstacle to this strategy. The challenges in analyzing the behavior are rooted in a lack of understanding about the nature of the return current. While local neutralization of the transmitting and receiving spacecraft is discussed below, we do not expect that an electron beam leaving the solar system will change the net charge of the solar system; a return current must form. Due to the very low density of the background plasma, that return current must occupy a large volume of space. If that return current takes the form of a cylindrical sheath surrounding the beam, the beam plus return current system will be *magnetically neutralized* and not respond to large-scale magnetic fields. Another way of saying this is that the beam and the background space plasma which carry the interplanetary or interstellar magnetic fields both have high magnetic Reynolds numbers, suggesting that the magnetic field near the beam should be dominated by the self-field of the beam, and the magnetic field far from the beam should be dominated by the *frozen* in fields of the background plasma, without much mixing. However, the return current of the interplanetary current sheet does not behave this way; instead of flowing back along the sheet, it spreads out in the termination shock and heliopause regions and eventually returns in a diffuse stream along the solar poles.   If the return current path of the beam were to follow a similar path it would not be magnetically neutralized and would be bent by the interplanetary magnetic field.

If the beam were bent by the interplanetary field, the radius of curvature, at $\gamma$ of 100, would be on the rough order of $10^{-3}$ AU.  While that might or might not be an insuperable barrier to this beam strategy, it would certainly be a considerable complication.  Further study of the interaction of the beam magnetic self-field and the "frozen-in" background plasma magnetic field will require particle in cell (PIC) [31] or similar simulation to resolve.

While this is only a preliminary study, it appears that the propagation of relativistic electron beams, via relativistic confinement, pinched propagation through space plasmas, or a combination of the two may be possible over ranges inaccessible to other techniques (hundreds to thousands of AU), even at GW or higher power levels.  This poses challenges in both beam generation and beam reception; however, if those could be overcome, it offers a potential route towards more practical interstellar beam-driven acceleration.  One purpose of this paper is to draw attention to the pieces missing.

## 3. Sources of Energy for Beamed Propulsion

Gigawatt class beams are a tremendous challenge for space-based systems.   Near-earth solar collection systems offer hundreds of watts per square meter – even at a very aggressive 500 W/m$^2$, a gigawatt scale system would require $2 \times 10^6$ m$^2$, a collection radius of 800 m.  This is comparable to the large satellites which have been proposed [32] [26] for space-based power collection for transmission to Earth.   This remains a credible option but obviously requires a degree of space-based infrastructure which we cannot yet estimate the schedule for. An Earth-orbiting asset could also be powered by energy beamed up from the surface of the Earth using similar microwave techniques to those envisioned for beaming solar power down. Both options introduce complications in that directing the beam to the accelerating spacecraft with very high pointing accuracy will require moving the beam along the predicted path of the spacecraft (the velocities being great enough that closed-loop guidance is very challenging).   Both also involve substantial pieces of in-space infrastructure which will carry a significant capital cost.

Another option that is opened up by beam ranges very large compared to an AU is to move the transmitter closer to the sun.   One of the main drivers of the low specific power (power per unit mass) of solar collection systems is the need to reject waste heat at something resembling Earth surface temperature (current photovoltaics, for example, lose efficiency much above 320 K).  A power collection and beam conversion system which operates at a much higher heat rejection temperature could operate with higher solar flux.

A suitable conversion approach is thermionic conversion; electrons emitted from a hot cathode and collected across a vacuum or low-pressure vapor gap on a cooler anode [33] [34]. The approach lends itself to operation at high solar flux in several ways – first, because it can operate at high anode temperatures, especially if there is also a difference in the work function of the materials employed in the cathode and the anode.   Secondly, because such devices can be, and have been, made with thin films for both materials [35], [36], [37]. Thirdly, because the thermionic emission from the cathode effectively cools the cathode, allowing (indeed, requiring) a high heat flux per unit area.

The effective power density of thermionic devices has ranged in the 50-100 kW/m² range [35] [36]. The limiting efficiency of a thermionic converter is essentially the Carnot efficiency (it is a heat engine using the electrons as the working fluid). At 0.05 AU, with a hot-side temperature of 2000 K and a cold-side temperature of 1300 K, with waste heat radiating from the two sides to space, the solar heat will drive 50 kW/m² of useful heat extracted with 26.2% of Carnot-limited efficiency. With 0.5 mil foil thickness for tungsten cathode and cobalt anode, plus an allowance for thin, lightweight spacers deposited on one of the films before assembly [37], areal density would be approximately 0.35 kg/m² (shown in Figure 2). 1 GW of useful power output at 50 kW/m² is an area of 20,000 m² or, as a circle, a radius of 80 m, and a mass (ignoring deployment structures or transmitter hardware) of 7000 kg. While those other components will certainly add to the mass, this is getting into the range where a launch or two of commercial launch vehicles plus an electric propulsion system for a slow cruise to the near-solar station might actually be realizable with a credible effort.

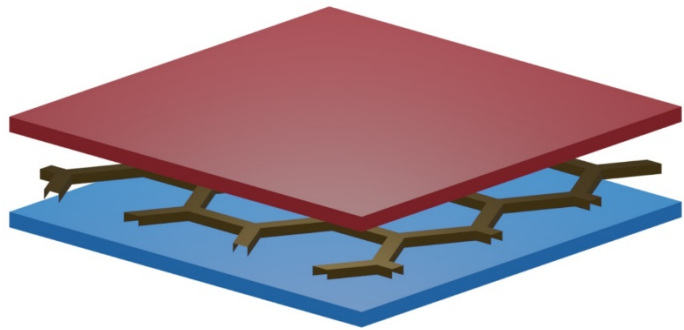

**Exploded view of thermionic stack**
Cathode in red, patterned spacer in gold, anode in blue

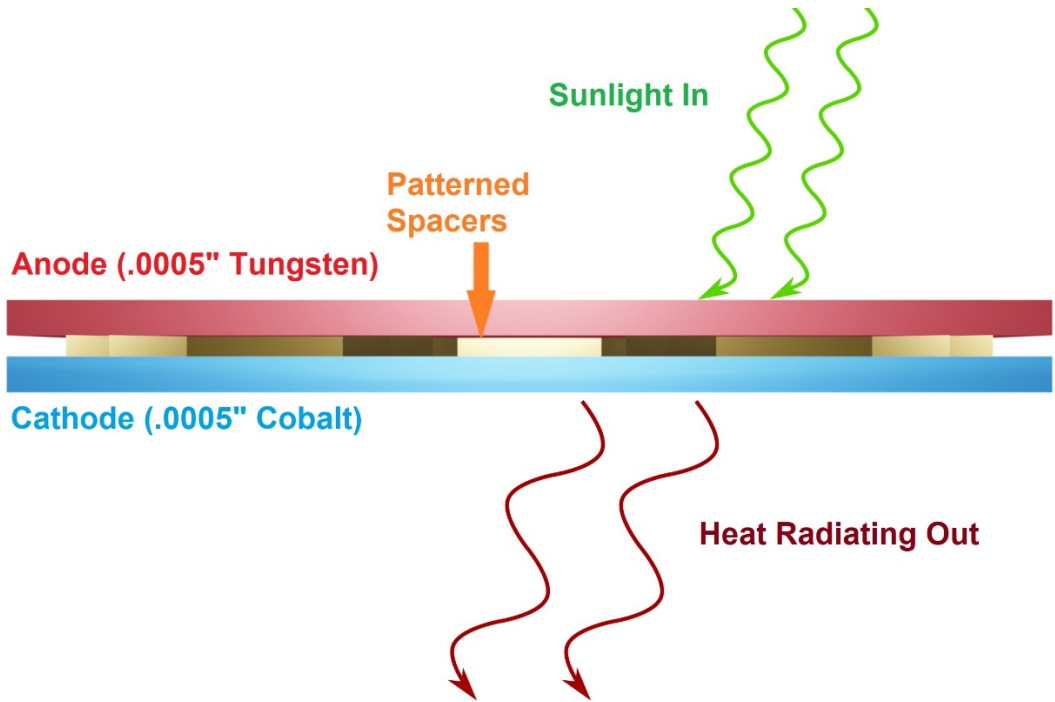

*Figure 2: Exploded and cross-section view of thermionic power collection.*

However, a solar orbit that close has a period of only four days; if we want to stay in the beam for weeks and months, this is an obvious problem for destinations close to the ecliptic. Some destinations, such as Alpha Centauri, are well out of the plane of the ecliptic and it is possible that a solar orbiting option exists. However, a potentially more attractive option is a solar statite. That would enable the transmitter to stay in a fixed orientation, with the hot side facing the Sun, the cold side and all of the beam transmitting hardware facing the target star system, and all of the hardware on the cold side to stay shadowed by the thermionic collection array. These distances are comparable to the perihelion of the Parker Solar probe, so keeping all of the instrumentation on the shadowed side of the collector is likely to be important.

While uncertainties in the structure of the solar wind remain and are currently being explored by the Parker Solar Probe, current models of the solar wind suggest that by 0.04 AU solar wind velocity is at least 300 km/s [20]. Therefore, in principle, a magnetic sail or plasma magnet can provide the force required to "hover" over the sun. Light pressure will not do it; light pressure can only provide hovering statite action for objects with areal density below roughly $10^{-3}$ kg/m$^2$, far below the expected density for the thermionic converter. Even allowing for an overall mass in place of 20,000 kg, at 0.05 AU, hovering thrust is 48,000 N. At the particle density from Table 1, dynamic pressure of the solar wind for the statite should be $\approx 5 \times 10^{-6}$ Pa at the lower-bound velocity estimate of 300 km/s, requiring a statite to intercept an area of solar wind $\approx 10^{10}$ m$^2$. Assuming solar wind dynamic pressure also scales inverse-square with distance from the Sun, the intercepting area would need to be 3x10$^{10}$ m$^2$; that corresponds to a radius of action of the plasma magnet of roughly 50 km. Given the plasma magnet performance of Slough [38] [39], that is an achievable level of solar wind drag. Additionally, given a gain between excitation of the rotating magnetic field for a plasma magnet and drag of 5000-10,000 [40], the hotel power for the hover should be 1.5-3 MW, or less than 0.3% of the thermionic power. An example of a possible configuration for the statite is in Figure 3 below.

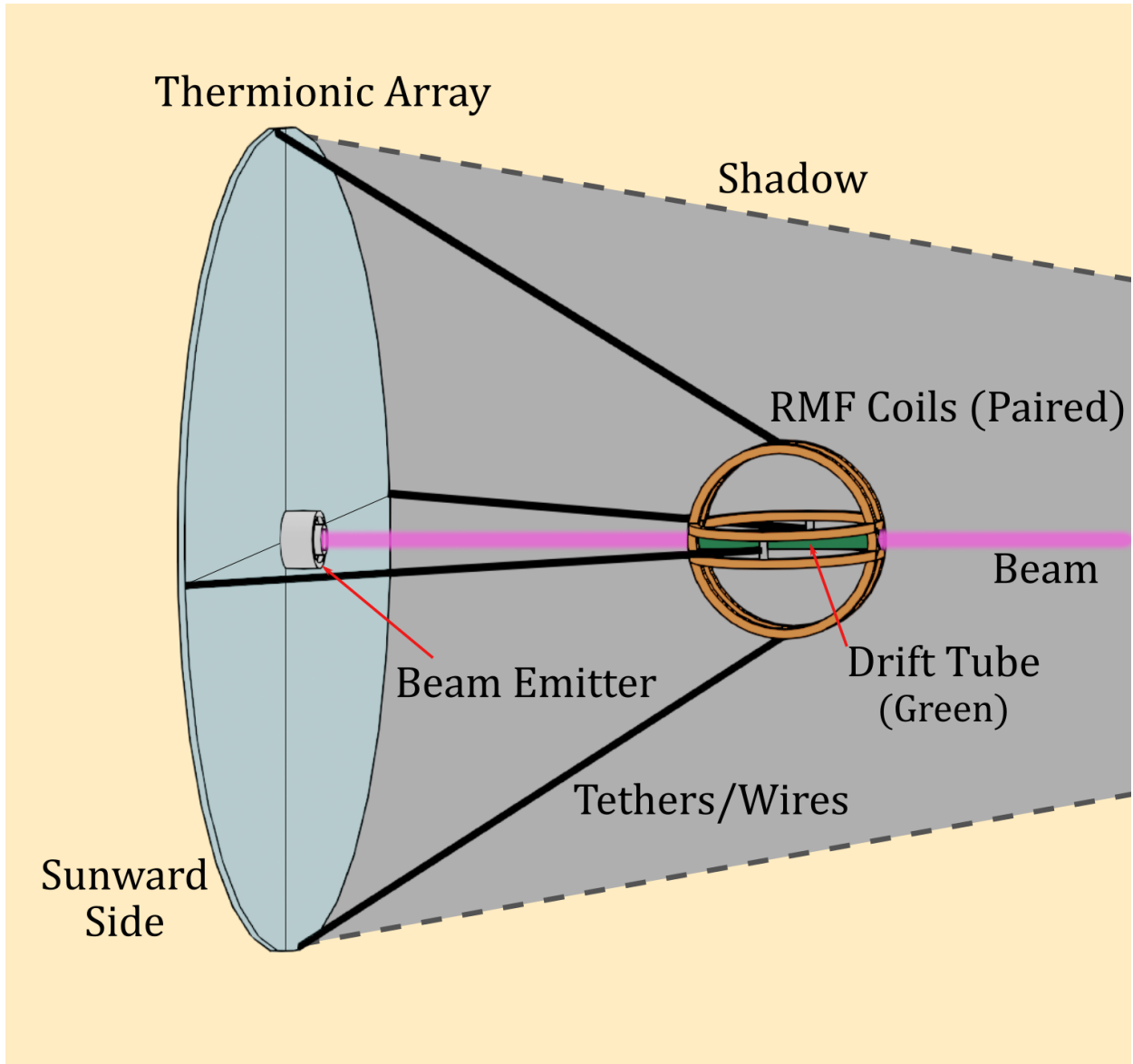

*Figure 3: Example configuration of thermionic power and electron beam transmission statite.*

## 4. Beam transmission

Given that the entire power output of the thermionic converter proposed above is in the form of flowing electrons, tapping off a fraction of those electrons to accelerate to the beam voltage is straightforward and similar to the commonly used thermionic electron sources for particle accelerators [41]. However, the charge must still be neutralized, or the transmitting spacecraft will build up an unacceptably positive potential. The return current, even if it flows back in a cylinder surrounding the electron beam, is too diffuse to connect with solid electrodes of a practical size, so the transmitting spacecraft will have to send out a stream of ions to neutralize the spacecraft (in terms of looking at the space plasma return current, these ions serve as a plasma contactor). Even at the lower range of $\gamma$ for our reference case, a 1 GW beam has a current of 20 A. A 0.1-year-long beam transmission then consists of $4 \times 10^{26}$ electrons. If the spacecraft carries a supply of lithium and singly ionizes a lithium atom for each emitted electron, that is ≈0.5 MJ/mole of ionization energy, and since each mole then yields $6.02 \times 10^{23}$ electrons, only 4.6 kg of lithium is required; the spacecraft can easily carry enough neutralization material for a long lifetime of operation. Power required for ionizing the neutralizing ions, ideally, is 100 W and quite negligible. Higher $\gamma$ operation as required for relativistic confinement without relying on pinch effects only reduces the beam current and neutralization requirements. The collection area to recover solar wind electrons sufficient for this purpose would mass more than the sacrificial neutralization ion system.

Even for quasi-static low frequency modulated voltages (for *bunching* the electrons), vacuum breakdown limits voltages to ~10 MV/m. If high frequency RF acceleration is used, electron accelerators now approach 100 MV/m [42] with >200 MV/m being explored using cryogenic techniques [43]. While it would involve a long tether, even the quasi-static acceleration is plausible; the plasma magnet array could be anti-sunward on a long tether, the thermionic array hanging from it under solar gravity. However, electrons are currently being accelerated in more compact plasma accelerators with accelerating potential up to 100 GV/m; if those techniques can be applied here, less than a meter of acceleration distance is required even for the purely relativistic confinement case [44], [45]. Whether that is lighter than the tethered method is unclear.

The physics of starting the beam and initially directing it to the receiver target have not been treated here. The physics at the beam head is more complex and will require more analysis. The challenge in designing the beam head is to avoid instabilities during the process of expelling the background plasma electrons to enter the ion-focused regime. During that transition from the background to the steady state, the hose instability can arise, and the transient analysis of the beam head behavior remains as future work. However, since we are only speaking of a beam reaching a cooperative target, it is possible that extremely fine pointing of the beam may be replaced by charging the receiving spacecraft positively via use of an electron gun so as to draw in the beam (analogous to striking an arc in denser gases). The beam can use whatever pulse shape, current, and repetition rate are most desirable for starting the beam; later pulses, once the beam is established, will tend to guide down the channel of ions already formed by the beam head expelling the background electrons.

## 5. Methods of Charged-Particle Beam Reception

The problem of receiving the beam and converting it to power is a difficult and important one which we currently consider unsolved. However, there are some possible avenues of approach:

5.1) *Electromagnetic Reflection*

For more conventional charged particle streams, the conventional method of receiving the beam would be to reflect them from a magnetic sail. Such an approach has the advantage of not heating the receiver on the ship and hence allowing for very high specific power. In the case of the relativistic electron beam, this essentially reduces the thrust per unit power to the same as would be experienced by a photon beam, which is wasteful of power for a target being accelerated to, for example, 0.1-0.2*c*. However, those drawbacks could be accepted; the more challenging problem is that the high magnetic rigidity of the electron beams that may be considered. For purely relativistic confinement of the beam, one may consider a $\gamma$ of 36500 which corresponds to an electron beam energy of ≈19 GeV. Such an electron beam has a magnetic rigidity of ≈62 T-m, which will require a very strong magnetic field to deflect. The high $\gamma$ beam also must be deflected gradually over a large volume to prevent excessive synchrotron radiation losses [15]. Conversely, a beam that relies entirely on pinch effects for confinement may consider beams with $\gamma$ as low as 100 ($\approx$50 MeV electron energy). These beams are much more easily deflected and could be good candidates for purely electromagnetic reflection.

5.2) *Direct Non-Thermal Rocket Conversion*

Since almost all the energy in the beam is in the kinetic energy of the particles rather than their rest mass, the beam is most effectively used as a source of energy, rather than a source of momentum. Of course, at these energies, a non-thermal means of transferring the energy into the exhaust velocity of reaction mass would be required. Classically, a rocket that minimizes energy use should have exhaust velocity of 0.65 of the delta-velocity for the mission. Thus, a ship being accelerated to 0.1$c$ should ideally have an exhaust velocity of $\approx$0.065$c$ (2x10$^7$ m/s).

This is the area where the greatest development is required before a credible beam-accelerated spacecraft using this strategy would be achievable. Currently, the authors do not have a satisfactory solution. However, it seems likely that the strategy involves coupling the electron beam motion into a plasma. Wake-field accelerators transfer energy from laser or particle beams passing through a plasma into high intensity electric fields in the plasma, which in turn accelerate electrons [44], [45] with accelerating potential up to $\approx$100 GV/m. What is needed for the non-thermal electron-beam-driven rocket is for the electron beam to excite high amplitude fields in the plasma inside a suitable waveguide cavity, which in turn couples to a backwards plasma wave [46], [47]. Backwards wave oscillators, especially plasma filled versions [48], as well as relativistic klystrons [49], may provide a path towards effective beam-plasma coupling. The plasma wave carries momentum, and that will push the plasma out the back (nozzle) end of the waveguide.

For a 1 GW beam with 100 AU range, the average specific power during the acceleration phase of $\approx$600 MW/kg is required to reach 0.1$c$. Using the scaling parameters above, with a 2.5-fold increase in electron energy for relativistic confinement, or use of pinched beams, the range could be increased ten-fold, still requiring $\approx$60 MW/kg. Clearly, for this to be practical, essentially all of the input energy has to be coupled into the exhaust or pass out of the device; similar to a laser sail. If any more than a small fraction of the throughput power is deposited into the device, the device temperature would rise to intolerable levels and require extensive cooling equipment. Furthermore, the electron beam cross section is reasonably large (tens of meters), and it would be very desirable for the reception process to include some focusing of the electron beam energy via electric or magnetic fields (such as a long tether with a positive charge extending along the beam axis).

## 6. Conclusions

By using relativistic electron beams, it is possible to transport substantial energy over distances that are orders of magnitude further than is practical with comparable infrastructure cost for electromagnetic beams. The power required to achieve interstellar-relevant transportation velocities is directly related to the distance over what that power is applied; increasing the distance could potentially greatly expand the usefulness of beam-powered interstellar spacecraft.

By using thermoelectric conversion in a near-solar statite, large, GW-class beam infrastructure can potentially be launched in the near term without waiting for the industrialization of near-Earth space. However, the usefulness of this approach depends on the discovery of a high-specific-power means of reception and conversion of that energy, either as beamed momentum or to eject reaction mass or push on the interplanetary medium as reaction mass.   While we have pointed out a possible avenue of approach, we do not yet have a satisfactory solution to that problem.

## 7. Further Work

First and foremost, the plasma physics of backwards-wave plasma oscillators excited by the relativistic electron beam should be explored to see if a compact package without much waste heat deposited in the structure is possible. Launching those waves in the external space plasma could equally provide useful propulsive means by propeller-like rather than rocket-like mechanisms.

Should such an approach be found, further development on the electron beam launching and thermionic converter system is called for.   We also believe that heavy ion beams such as lead nuclei in a self-pinched strategy needs further exploration. The electron beam physics require further study, especially in the area of the beam head and in the interaction with magnetic fields frozen in to the background plasma. These capabilities could be useful, over shorter ranges and lower power levels, for fast transits inside the solar system; such missions could be useful precursors to develop the technology and are worth study.

## 8. Funding

## 9. Acknowledgement

Thanks to the ToughSF Discord group, which brought the two authors together to explore these topics.  Thanks to Max Greason who provided some of the figures.   We are grateful to the reviewers at Acta Astronautica who provided detailed comments which materially improved the paper.

## 10. Vitae

*Jeffrey K. Greason*

Jeffrey Greason is the Chief Technologist of Electric Sky, developing long-range wireless power for propulsion and other purposes; and Chairman of the Tau Zero Foundation, developing advanced propulsion technologies for solar system and interstellar missions.  He has been active in the development of commercial space regulation, served on COMSTAC, helped craft the Commercial Space Launch Amendments Act of 2004, and served on the "Review of U.S. Human Space Flight Plans Committee" in 2009.  He was a cofounder of XCOR Aerospace and served as CEO from 1999 to early 2015, developing numerous reusable rocket engines and two rocket powered aircraft with 66 successful test flights.  During that time, he co-founded the industry trade association for commercial space companies, the Commercial Spaceflight Federation.  Previously, he was the rocket engine team lead at Rotary Rocket, and an engineering manager in chip technology development at Intel.  He holds 29 U.S. Patents and has recently published several papers on novel space propulsion concepts.  He is an

Associate Fellow of the AIAA and a member of the Nuclear and Future Flight propulsion technical committee.

*Gerrit Bruhaug*

Gerrit Bruhaug is a Harold Agnew National Security Fellow at Los Alamos National Laboratory where he works on advanced radiation sources from high-intensity lasers and particle accelerators.